\documentclass{aa}
\usepackage{graphicx}
\usepackage{txfonts}
\def\Cau  {Ca\,{\small I}}
\def\Cad  {Ca\,{\small II}}
\def\Cat  {Ca\,{\small III}}
\def\Feu  {Fe\,{\small I}}

\def\Teff  {$\rm T_{eff}$}
\def\logg  {$\log g$}
\def\kms   {$\rm km s^{-1}$}
\begin{document}

\title{
NLTE determination of the calcium abundance and 3D corrections in
extremely metal-poor stars
\thanks{Based on observations obtained with the ESO Very Large
Telescope at Paranal Observatory, Chile (Large Programme ``First
Stars'', ID 165.N-0276(A); P.I.: R. Cayrel).}
\thanks{The NLTE corrections of the Ca lines are available in electronic form
at the CDS via anonymous ftp to cdsarc.u-strasbg.fr (130.79.128.5)
or via http://cdsweb.u-strasbg.fr/cgi-bin/qcat?J/A+A/}
}

\author {
M.~Spite\inst{1}\and
S.M.~Andrievsky\inst{1,2}\and
F.~Spite\inst{1}\and
E.~Caffau\inst{3,1}\and
S.A.~Korotin \inst{2}\and
P.~Bonifacio\inst{1}\and
H.-G.~Ludwig\inst{3}\and
P.~Fran\c cois\inst{1}\and
R.~Cayrel\inst{1}
}
\offprints{M. Spite\\
           e-mail: Monique.Spite@obspm.fr}
\institute {
GEPI Observatoire de Paris, CNRS, Universit\'e Paris Diderot, F-92195
Meudon Cedex France e-mail : {\tt monique.spite@obspm.fr}
\and  
Department of Astronomy and Astronomical Observatory, Odessa National
University, T.G. Shevchenko Park, 65014, Odessa, Ukraine, and Isaac
Newton Institute
of Chile, Odessa Branch, Ukraine
\and  
Zentrum f\"ur Astronomie der Universit\"at Heidelberg,
Landessternwarte, K\"onigstuhl 12, 69117 Heidelberg, Germany
}

\date{}

\authorrunning{Spite et al.}
\titlerunning{NLTE determination of the abundance of Calcium in EMP 
stars}

  \abstract
   {Calcium is a key element for constraining the models of chemical
   enrichment of the Galaxy.}
   {Extremely metal-poor stars contain the fossil records of the chemical 
   composition of the early Galaxy and it is important to compare Ca 
   abundance with abundances of other light elements, that are supposed 
   to be synthesized in the same stellar evolution phases.  }
   {The NLTE profiles of the calcium lines were computed in a
   sample of 53 extremely metal-poor stars with a modified version of
   the program MULTI, which allows a very good description of the
   radiation field.  }
   {With our new model atom we are able to  reconcile the abundance 
   of Ca deduced from the \Cau~ and \Cad~ lines in Procyon. This 
   abundance is found to be solar.\\ 
   --We find that $\rm \overline{[Ca/Fe]}=0.50 \pm 0.09$
   in the early Galaxy, a value slightly higher than the previous LTE
   estimations.\\ 
   --The scatter of the ratios
   [X/Ca] is generally smaller than the scatter of the ratio [X/Mg]
   where X is a ``light metal'' (O, Na, Mg, Al, S, and K) with the
   exception of Al.  These scatters cannot be explained by error of
   measurements, except for oxygen.  Surprisingly, the scatter of
   [X/Fe] is always equal to, or even smaller than, the scatter around the
   mean value of [X/Ca].\\
   --We note that at low metallicity, the wavelength of the \Cau~ 
   resonance line is shifted relative to the (weaker) subordinate lines, 
   a signature of the effect of convection.\\
   --The Ca abundance deduced from  the \Cau~ resonance line
   (422.7~nm) is found to be systematically smaller at very low metallicity,
   than the abundance deduced 
   from the subordinate lines. Our computations of the effects of  
   convection (3D effects) are not able to explain this difference.
   A fully consistent 3D NLTE model atmosphere and line
   formation scheme would be necessary to fully capture the physics of the
   stellar atmosphere.  
   }
   {}

\keywords{ Line : Formation -- Line : Profiles -- Stars: Abundances --
Stars: Supernovae -- Galaxy evolution}

\maketitle

\section{Introduction}
An homogeneous sample of 53 metal-poor stars, most of them extremely 
metal-poor (EMP stars with $\rm [Fe/H]<-2.9$), has been observed 
by Cayrel et al.  (\cite{CDS04}), and Bonifacio et al.  (\cite{BMS07}, \cite{BSC09}).  
The aim of this paper is to determine more precisely the calcium abundance in these stars.
 120-125  F6.2  ---     Corr18    NLTE corrections, model (7000,2,-3.5,0.3)
These low mass stars have been formed at the very early phases of the Galaxy and the chemical composition of their atmosphere reflects the yields of the first massive type II supernovae which have a very short life-time.  
These supernovae produce more ``$\alpha$-elements'' (O, Mg, Si, S, Ca)
than ``iron-peak'' elements.
In contrast, less massive type I supernovae, which have a much longer life-time and explode later, produce more iron-peak
elements than $\alpha$-elements.  As a consequence in the atmosphere of the EMP stars formed at the beginning of the Galaxy
a relative enhancement of the $\alpha$-elements (compared to the Sun) is observed.  The level of this overabundance is one of the fundamental parameters of the chemical evolution models of the Galaxy.

On the other hand, the relative production of the $\alpha$-elements in
a supernova depends on the mass of this supernova (e.g. Kobayashi et
al., \cite{KUN06}): [Ca/Fe] tends to be lower for more massive
supernovae.  Therefore the level of [Ca/Fe] in the early Galaxy constrains
the IMF in the early times.

Moreover, Cayrel et al. (\cite{CDS04}) and Bonifacio et
al.  (\cite{BSC09}) have shown that for $\rm[Fe/H]<-2.7$ the
abundances of the $\alpha$-elements relative to iron ([Mg/Fe],
[Si/Fe], [Ca/Fe], and [Ti/Fe]) are constant with a small scatter but
that surprisingly the scatters of these elements relative to magnesium
(another $\alpha$-element) are larger.  This larger scatter cannot be
explained by a lower precision of the abundance of magnesium.
Cayrel et al.  (\cite{CDS04}) and Bonifacio et al.  (\cite{BSC09}) computed the
abundances under the LTE hypothesis.  The non-LTE (NLTE)
abundances of two $\alpha$-elements Mg and S have been then computed in
Andrievsky et al.  (\cite{ASK10}) and Spite et al.  (\cite{SCA11}).
These papers confirm that the scatter of abundance ratios is
generally larger when Mg replaces Fe as a reference element. 120-125  F6.2  ---     Corr18    NLTE corrections, model (7000,2,-3.5,0.3)
It is then interesting to check wether the link between the abundance
of Ca and the abundance of the other $\alpha$-elements is closer.

In this paper we have carried out a NLTE analysis of the calcium abundance in the atmosphere of these EMP stars and have tried to estimate the influence of convection (3D computations).  

For this new analysis we have used the resonance line of \Cau~ and about 15 subordinate lines. We have also used the line of the \Cad~ infrared triplet at 866.21nm.
Unfortunately, the two other lines of this triplet (at 849.81 and
854.21nm) are outside the observed spectral range.

\begin{table*}
\begin{center}    
\caption[]{ Program stars and their parameters. 
The stars marked with an asterisk are carbon-rich.
Columns 2 to 5 give the main parameters of the
stars.  Column
6 is the NLTE calcium abundance deduced from the subordinate \Cau~
lines with the number of lines and the standard deviation (columns 7
and 8).  Columns 9 and 10 list the abundance of Ca derived from a
NLTE computation of the 422.67 and 866.22nm lines.  The last two
columns give [Ca/H] and [Ca/Fe] based on the calcium abundance deduced
from the \Cau~ subordinate lines.  
}
\label{tabstars}
\begin{tabular}{lcccccccccrrr}
~~~1  &     2         &     3    &    4         &   5   &    6               &  7&    8    &         9           &       10~~~       &     11~~~~&  12~~~~~\\
\hline              
~~~star&$T_{\rm eff}$ & $\log~g$ & $\xi_{\rm t}$  &[Fe/H] & $\log\epsilon$(Ca) & N &$\sigma$ &$\log \epsilon$(Ca)  & $\log\epsilon$(Ca) &[Ca/H] &[Ca/Fe] \\
      &     K         &          &  km~s$^{-1}$ &       &  sub. lines&   &         &    422.67~nm    &   866.22~nm     &       &        \\

\hline
{\bf turnoff stars} \\
\hline
BS~16023--046  &   6360&  4.5&  1.4&     --2.97&      3.77&  8& 0.09& 3.54 &     3.74 &    --2.59 &     0.38\\
BS~16968--061  &   6040&  3.8&  1.5&     --3.05&      3.77& 11& 0.08& 3.70 &     3.70 &    --2.59 &     0.46\\
BS~17570--063  &   6240&  4.8&  0.5&     --2.92&      3.80& 12& 0.06& 3.70 &     3.70 &    --2.56 &     0.36\\
CS~22177--009  &   6260&  4.5&  1.2&     --3.10&      3.64& 11& 0.09& 3.52 &     3.61 &    --2.72 &     0.38\\
CS~22888--031  &   6150&  5.0&  0.5&     --3.28&      3.45& 10& 0.11& 3.33 &     3.36 &    --2.91 &     0.37\\
CS~22948--093  &   6360&  4.3&  1.2&     --3.43&      3.49&  4& 0.06& 3.37 &     3.59 &    --2.87 &     0.56\\
CS~22953--037  &   6360&  4.3&  1.4&     --2.89&      3.83& 11& 0.10& 3.71 &     3.73 &    --2.53 &     0.36\\
CS~22965--054  &   6090&  3.8&  1.4&     --3.04&      3.93& 12& 0.22& 3.78 &     3.71 &    --2.43 &     0.61\\
CS~22966--011  &   6200&  4.8&  1.1&     --3.07&      3.69& 11& 0.10& 3.60 &     3.51 &    --2.67 &     0.40\\
CS~29499--060  &   6320&  4.0&  1.5&     --2.70&      4.05& 14& 0.12& 3.94 &     4.08 &    --2.31 &     0.39\\
CS~29506--007  &   6270&  4.0&  1.7&     --2.91&      4.05& 14& 0.09& 3.95 &     3.85 &    --2.31 &     0.60\\
CS~29506--090  &   6300&  4.3&  1.4&     --2.83&      4.09& 14& 0.11& 3.90 &     3.93 &    --2.27 &     0.56\\
CS~29518--020  &   6240&  4.5&  1.7&     --2.77&      4.03& 10& 0.18& ~~-- &     ~~-- &    --2.33 &     0.44\\
CS~29518--043  &   6430&  4.3&  1.3&     --3.24&      3.71& 11& 0.17& 3.54 &     3.60 &    --2.65 &     0.59\\
CS~29527--015  &   6240&  4.0&  1.6&     --3.55&      3.35&  4& 0.20& 3.06 &     3.26 &    --3.01 &     0.54\\
CS~30301--024  &   6330&  4.0&  1.6&     --2.75&      4.16& 15& 0.08& 4.00 &     4.04 &    --2.20 &     0.55\\
CS~30339--069  &   6240&  4.0&  1.3&     --3.08&      3.81& 11& 0.13& 3.79 &     3.77 &    --2.55 &     0.53\\
CS~31061--032  &   6410&  4.3&  1.4&     --2.58&      4.23& 14& 0.14& 4.16 &     ~~-- &    --2.13 &     0.45\\
\hline
{\bf giants}  \\
\hline
HD~2796        &   4950&  1.5&  2.1&     --2.47&      4.33& 16& 0.09& 4.32 &     4.45 &    --2.03 &     0.43\\
HD~122563      &   4600&  1.1&  2.0&     --2.82&      3.97& 16& 0.08& 3.93 &     3.95 &    --2.39 &     0.42\\
HD~186478      &   4700&  1.3&  2.0&     --2.59&      4.31& 16& 0.11& 4.33 &     4.12 &    --2.05 &     0.53\\
BD~+17$^{\circ} $3248& 5250& 1.4& 1.5&   --2.07&      4.74& 14& 0.05& 4.74 &     4.70 &    --1.62 &     0.45\\
BD~--18$^{\circ}$5550& 4750& 1.4& 1.8&   --3.06&      3.88& 16& 0.09& 3.72 &     3.80 &    --2.48 &     0.58\\
CD~--38$^{\circ}$245 & 4800& 1.5& 2.2&   --4.19&      2.72&  4& 0.13& 2.30 &     2.64 &    --3.64 &     0.55\\
BS~16467--062  &   5200&  2.5&  1.6&     --3.77&      3.22&  9& 0.23& 2.97 &     2.97 &    --3.14 &     0.63\\
BS~16477--003  &   4900&  1.7&  1.8&     --3.36&      3.59& 16& 0.12& 3.32 &     3.45 &    --2.77 &     0.59\\
BS~17569--049  &   4700&  1.2&  1.9&     --2.88&      4.04& 16& 0.14& 4.06 &     3.90 &    --2.32 &     0.56\\
CS~22169--035  &   4700&  1.2&  2.2&     --3.04&      3.64& 16& 0.09& 3.34 &     3.70 &    --2.72 &     0.32\\
CS~22172--002  &   4800&  1.3&  2.2&     --3.86&      3.11& 11& 0.09& 2.52 &     3.11 &    --3.25 &     0.61\\
CS~22186--025  &   4900&  1.5&  2.0&     --3.00&      3.89& 15& 0.08& 3.73 &     3.90 &    --2.47 &     0.53\\
CS~22189--009  &   4900&  1.7&  1.9&     --3.49&      3.27& 12& 0.07& 2.95 &     3.40 &    --3.09 &     0.40\\
CS~22873--055  &   4550&  0.7&  2.2&     --2.99&      3.89& 16& 0.09& 3.73 &     3.75 &    --2.47 &     0.52\\
CS~22873--166  &   4550&  0.9&  2.1&     --2.97&      3.92& 16& 0.09& 3.84 &     3.92 &    --2.44 &     0.53\\
CS~22878--101  &   4800&  1.3&  2.0&     --3.25&      3.68& 16& 0.12& 3.30 &     3.65 &    --2.68 &     0.57\\
CS~22885--096  &   5050&  2.6&  1.8&     --3.78&      3.10&  8& 0.11& 2.83 &     2.98 &    --3.26 &     0.52\\
CS~22891--209  &   4700&  1.0&  2.1&     --3.29&      3.58& 16& 0.08& 3.33 &     3.45 &    --2.78 &     0.51\\
CS~22892--052* &   4850&  1.6&  1.9&     --3.03&      3.84& 14& 0.12& 3.65 &     3.80 &    --2.52 &     0.51\\
CS~22896--154  &   5250&  2.7&  1.2&     --2.69&      4.12& 16& 0.12& 4.01 &     4.00 &    --2.23 &     0.46\\
CS~22897--008  &   4900&  1.7&  2.0&     --3.41&      3.45& 11& 0.12& 3.05 &     3.35 &    --2.91 &     0.50\\
CS~22948--066  &   5100&  1.8&  2.0&     --3.14&      3.69& 14& 0.09& 3.45 &     3.60 &    --2.67 &     0.47\\
CS~22949--037* &   4900&  1.5&  1.8&     --3.97&      3.01&  9& 0.10& 2.75 &     3.20 &    --3.35 &     0.61\\
CS~22952--015  &   4800&  1.3&  2.1&     --3.43&      3.29& 12& 0.08& 2.95 &     3.25 &    --3.07 &     0.36\\
CS~22953--003  &   5100&  2.3&  1.7&     --2.84&      3.89& 16& 0.10& 3.74 &     3.85 &    --2.47 &     0.37\\
CS~22956--050  &   4900&  1.7&  1.8&     --3.33&      3.69& 16& 0.09& 3.49 &     3.75 &    --2.67 &     0.66\\
CS~22966--057  &   5300&  2.2&  1.4&     --2.62&      4.23& 16& 0.09& 4.21 &     4.10 &    --2.13 &     0.49\\
CS~22968--014  &   4850&  1.7&  1.9&     --3.56&      3.06& 10& 0.10& 2.67 &     2.89 &    --3.29 &     0.27\\
CS~29491--053  &   4700&  1.3&  2.0&     --3.04&      3.89& 16& 0.08& 3.68 &     3.85 &    --2.47 &     0.57\\
CS~29495--041  &   4800&  1.5&  1.8&     --2.82&      4.05& 16& 0.07& 4.00 &     3.95 &    --2.31 &     0.51\\
CS~29502--042  &   5100&  2.5&  1.5&     --3.19&      3.55& 13& 0.08& 3.41 &     3.40 &    --2.81 &     0.38\\
CS~29516--024  &   4650&  1.2&  1.7&     --3.06&      3.99& 16& 0.12& 3.93 &     ~~-- &    --2.37 &     0.69\\
CS~29518--051  &   5200&  2.6&  1.4&     --2.69&      4.14& 16& 0.07& 4.07 &     4.10 &    --2.22 &     0.47\\
CS~30325--094  &   4950&  2.0&  1.5&     --3.30&      3.61& 15& 0.08& 3.51 &     3.55 &    --2.75 &     0.55\\
CS~31082--001  &   4825&  1.5&  1.8&     --2.91&      4.01& 14& 0.08& 4.01 &     3.90 &    --2.35 &     0.56\\
\hline
\end{tabular}
\end{center}
\end{table*}

\section {Star sample and model parameters} 

The spectra of the stars investigated here have been presented in
detail in Cayrel et al.  (2004) and Bonifacio et al.  (\cite{BMS07}).
The observations were performed with the high-resolution spectrograph
UVES at ESO-VLT (Dekker et al., \cite{DDK00}). The resolving power of the spectrograph is $R\approx
45000$, with about five pixels per resolution element and the S/N ratio per
pixel is typically about 150.

The fundamental parameters of the models (effective temperature \Teff, logarithm of the gravity \logg, and metallicity)
have been derived by Cayrel et al.  (\cite{CDS04}) for the giants and
Bonifacio et al.  (\cite{BMS07}) for the turnoff stars.  Briefly,
temperatures of the giants are deduced from the colors with the
calibration of Alonso et al.  (\cite{AAM99}, \cite{AAM01}), and
temperatures of the turnoff stars from the wings of the $\rm
H{\alpha}$ line.  Moreover we checked that these $\rm
H{\alpha}$ temperatures agreed with the temperatures
derived from the color {\sl V-K} and the calibration of Alonso et al.
(\cite{AAM96}).  The gravities are derived from the ionization
equilibrium of iron (under the LTE approximation) and we note that
they might be affected by NLTE effects.  The parameters of the
models are repeated in Table \ref{tabstars} for the reader's
convenience. 

\section {Determination of the calcium abundance}

The NLTE profiles of the calcium lines were computed with a modified
version of the code MULTI (Carlsson, \cite{Car86}, Korotin et al.,
\cite{KAL99}), which allows a very good description of the radiation
field.  This version includes opacities from ATLAS9 (Kurucz,
\cite{Kur92}), which modify the intensity distribution in the UV. 

\subsection {Atmospheric models}  \label{smod}
For these computations we used Kurucz models without overshooting
(Castelli et al., 1997).  These models have been shown to provide LTE
abundances very similar (within 0.05~dex) to those of the MARCS models
used by Cayrel et al. (\cite{CDS04}) and Bonifacio et al.
(\cite{BSC09}).  The solar model was taken from Castelli\footnote
{http://wwwuser.oats.inaf.it/castelli/sun/ap00t5777g44377k1asp.dat}
with a chromospheric contribution from the \mbox{VAL-3C} model of 
Vernazza et al. (\cite {VAL81}) and the corresponding microturbulence 
distribution.


\subsection{Atomic model} \label{atomodel}

Our model atom of calcium is similar to the one used by
Mashonkina et al.  (\cite{MKP07}) but it includes some more levels and more recent atomic data. 
Seventy levels of \Cau, thirty-eight levels of \Cad, and the ground state of \Cat~ were taken into account; in addition, more than 300 levels of \Cau~ and
\Cad~ were included to keep the condition of the particle number
conservation in LTE. The fine structure was taken into account for the
levels $\rm 3d^{2}D$ and $\rm 4p^{2}P^{0}$ of \Cad.

The energy levels were taken from the NIST atomic spectra database
(Sugar \& Corliss, \cite{SC85}).  The ionization cross-sections
were taken from TOPBASE.
The oscillator strengths of the \Cau~ and \Cad~ lines were taken 
from the most recent estimations: 
Wiese et al. (\cite{WSM69}),
Smith \& Raggett (\cite{SR81}),
Smith (\cite{Smi88}),
Theodosiou (\cite{Theo89}),
Morton (\cite{Mor91}), 
Kurucz (\cite{Kur93}), and from TOPBASE for the lines occuring between non-splitted levels.
For the
forbidden transitions, we used TOPBASE and Hirata \& Horaguchi (\cite{HH95}).

We considered 351 transitions in detail; for 375
weak transitions the radiative rates were fixed.  Collisional rates
between the ground level and the ten lower levels of \Cau~ were taken
from Samson \& Berrington (\cite{SB01}).  For \Cad, collisional rates
were taken from Mel\'endez et al. (\cite{MBB07}) instead of those of
Burgess et al.  (\cite{BCT95}) used by Mashonkina et al.
(\cite{MKP07}) for the lower seven terms.

For the other transitions (without data) we used for allowed
transitions, the Van Regemorter (\cite{VR62}) formula, and for the
forbidden transitions the Allen (\cite{All73}) formula.

Electron impact ionization cross-sections were calculated by applying
the formula of Seaton (\cite{Sea62}), with threshold photoionization
cross-sections from the Opacity-Project data.

Collisions with hydrogen atoms were computed using the Steenbock \&
Holweger (\cite{SH84}) formula.  The cross sections calculated with
this formula were multiplied by a scaling factor $\rm S_{H}$: the
``efficiency'' of the hydrogenic collisions.  This factor was
constrained empirically by comparing the Ca abundance obtained from
different lines in the Sun and in some reference stars (Procyon,  HD\,140283, HD\,122563).  The best agreement was obtained
with $\rm S_{H}= 0.1$ which agrees well with Ivanova et al.
(\cite{ISS02}) and Mashonkina et al.  (\cite{MKP07}).

\begin{figure}[ht]
 \centering
  \includegraphics[width=0.36\textwidth,clip]{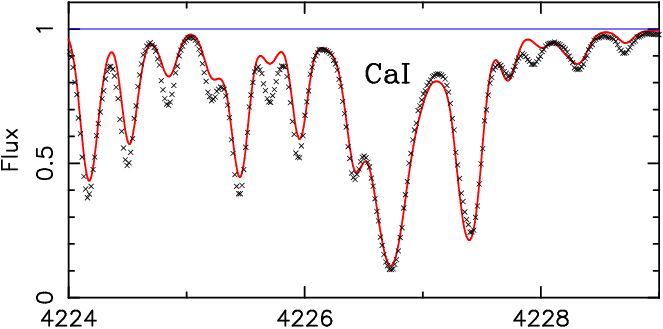}
  \includegraphics[width=0.36\textwidth,clip]{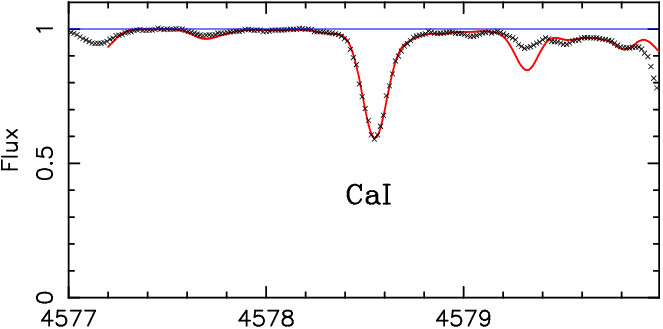}
  \includegraphics[width=0.36\textwidth,clip]{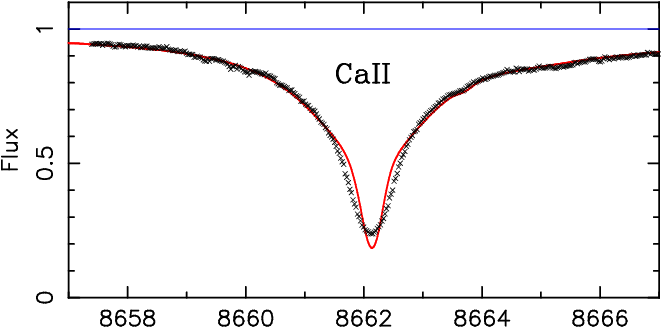}
  \includegraphics[width=0.36\textwidth,clip]{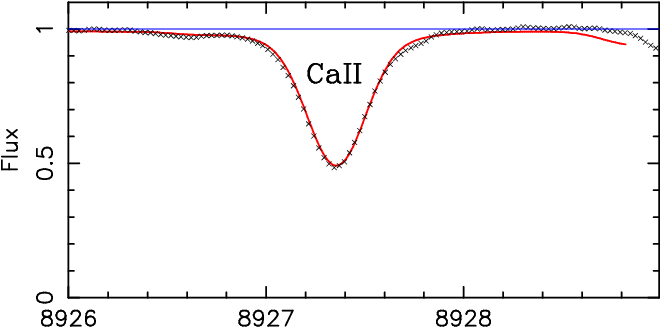}
\caption[]{Profiles of  Ca~I and Ca~II lines in Procyon. The wavelengths are in \AA. The small crosses represent the observed spectrum and the (red) line the computed profile. {\bf With our model atom, the observed and computed  profiles of the \Cau~  and \Cad~lines  agree well.}
}
\label{procyon}
\end{figure}

\begin{figure}[ht]
 \centering
  \includegraphics[width=0.36\textwidth,clip]{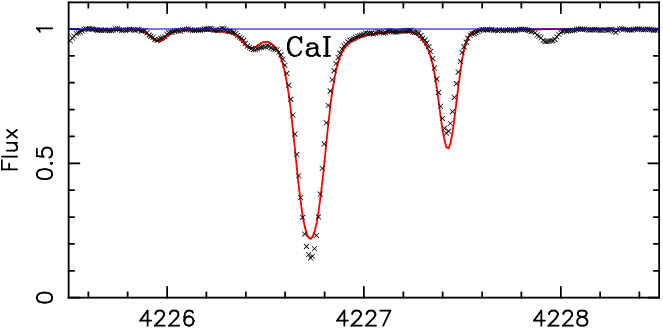}
  \includegraphics[width=0.36\textwidth,clip]{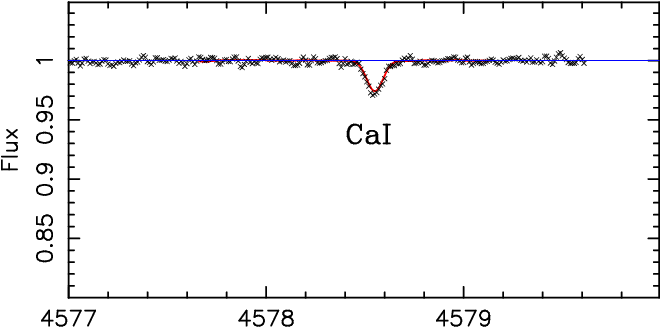}
  \includegraphics[width=0.36\textwidth,clip]{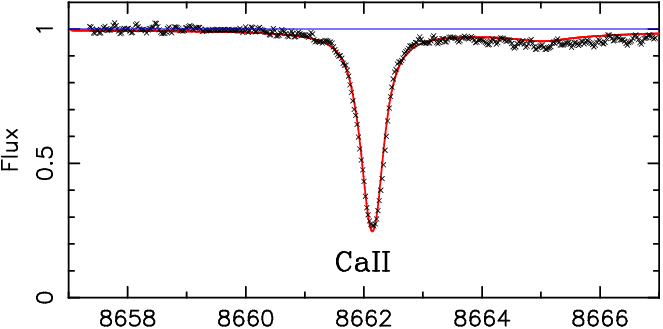}
  \includegraphics[width=0.36\textwidth,clip]{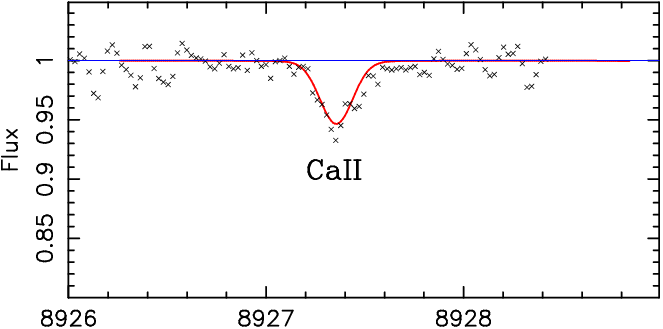}
     \caption[]{Profiles of  Ca~I and Ca~II lines in HD\,140283. The wavelengths are in \AA. The symbols are the same as in Fig. \ref{procyon}. The observed spectrum and the synthetic profiles computed with  $\log \epsilon$(Ca) =4.12 agree well.}
\label{HD140283}
\end{figure}

\subsection{Consistency check}
To test the Ca atom model, we computed the profiles of the calcium
lines in the Sun, in Procyon, and in two classical metal-poor stars  
HD\,122563, and HD\,140283. The synthetic spectra were computed following the procedure described in Korotin (\cite{Kor08}): we calculated the departure coefficients factors ``b'' for the Ca lines with MULTI and then used these factors in the LTE synthetic spectrum code.\\
 
$\bullet$ For the Sun we used the solar atmosphere model computed by Castelli   
 with a chromospheric contribution following Vernazza et al. (\cite{VAL81})  (see section \ref{smod}). A micro turbulence velocity 
 $\rm \xi_{t}=1.0$ \kms~ was adopted in the atmosphere.
 
We computed the profile of 49 lines of \Cau~ and 17 lines of \Cad. 
The $\log gf$ value of the lines and the broadening parameter due to collisions with hydrogen atoms,    $\log \gamma_{VW}/N_{H}$ (for T=10\,000K),  are given in Table \ref{sunlist}.
We found  $\log \epsilon$(Ca)  = $6.31 \pm 0.05$  from the \Cau~lines and 
$\log \epsilon$(Ca)  = $6.30 \pm 0.07$  from the \Cad~ lines. These values
agree well with the meteoritic calcium abundance  $\log \epsilon$(Ca)  = $6.31 \pm 0.02$ (Lodders et al., \cite{LPG09}) and the photospheric abundances derived by Asplund et al. (\cite{AGS09}) for the solar atmosphere:  $\log \epsilon$(Ca)  = $6.34 \pm 0.04$ (a value that includes 3D effects). 


For most of the lines in Table \ref{sunlist},  the broadening parameter $\log \gamma_{VW}/N_{H}$ (for T=10\,000K) due to collisions with hydrogen atoms, was taken from the precise calculations of Anstee \&  O'Mara (\cite{AO95}), Barklem \&  O'Mara (\cite{BO97}, \cite{BO98}), and Barklem et al. (\cite{BOR98}). For the other lines, this parameter was derived from the fit of the solar atlas (Kurucz et al., \cite{KFB84}). 
 These values are, for the high excitation lines of \Cad, higher than the values obtained from the Uns\"old  or Kurucz approximation. However, they agree well with the values obtained by Ivanova et al. (\cite{ISS02}), who note that the use of the 
Uns\"old  or Kurucz approximation often leads to an underestimation of the broadening parameter.\\ 

\begin{table}[ht] 
\begin{center}     
\caption[]{Parameters of the Ca lines used in the solar spectrum. 
$\gamma_{1}=\log \gamma_{VW}/N_{H}$ for a temperature of 10\,000K.
}
\label{sunlist} 
\begin{tabular}{c@{ }r@{ }c@{ }c@{ }c@{~~}c@{ }r@{ }c@{ }c@{ }c}
\hline 
line &	$\log gf$&    &	$\gamma_{1}$&   & line &	 $\log gf$&   &$\gamma_{1}$&   \\
(nm) &           & Ref&              &Ref&  (nm)&           &Ref&            &Ref\\
\hline
~~\Cau		&         &       &	  &   &~\Cau	\\ 
410.8526&  -0.824&  1&  -6.97&  11&  616.6439&  -1.143&  4&  -7.15&  9 \\ 
422.6728&   0.244&  6&  -7.56&   9&  616.9042&  -0.797&  4&  -7.15&  9 \\ 
428.3011&  -0.220&  3&  -7.50&  12&  616.9563&  -0.478&  4&  -7.15&  9 \\ 
428.9367&  -0.300&  3&  -7.50&  12&  643.9075&   0.390&  4&  -7.57&  9 \\ 
430.2528&   0.280&  3&  -7.50&  12&  645.5598&  -1.340&  2&  -7.65&  1 \\ 
431.8652&  -0.211&  3&  -7.50&  12&  646.2567&   0.262&  4&  -7.57&  9 \\ 
435.5079&  -0.420&  3&  -6.80&  11&  647.1662&  -0.686&  4&  -7.57&  9 \\ 
442.5437&  -0.360&  3&  -7.16&   9&  649.3781&  -0.109&  4&  -7.57&  9 \\ 
443.4957&  -0.010&  3&  -7.16&   9&  649.9650&  -0.818&  4&  -7.57&  9 \\ 
443.5679&  -0.523&  3&  -7.16&   9&  657.2779&  -4.296&  6&  -7.69&  9 \\ 
445.4779&   0.260&  3&  -7.16&   9&  671.7681&  -0.523&  4&  -7.14&  9 \\ 
445.6616&  -1.660&  3&  -7.16&   9&  714.8150&   0.137&  4&  -7.80&  1 \\ 
451.2268&  -1.892&  4&  -7.26&   1&  720.2200&  -0.262&  7&  -7.80&  1 \\ 
452.6928&  -0.548&  4&  -7.00&  11&  732.6145&  -0.208&  7&  -7.30& 11 \\ 
457.8551&  -0.697&  4&  -7.00&  11& 1034.3820&  -0.409&  5&  -7.12&  9 \\ 
468.5268&  -0.880&  3&  -6.40&  11&                                    \\ 
518.8844&  -0.115&  4&  -7.00&  11&  \Cad                              \\ 
526.0387&  -1.719&  4& \bf -7.51& \bf  9&  393.3663&   0.104&  6&  -7.76&  9 \\ 
526.1704&  -0.579&  4& \bf -7.51& \bf  9&  396.8469&  -0.201&  6&  -7.76&  9 \\ 
526.5556&  -0.114&  4& \bf -7.51& \bf  9&  500.1479&  -0.506& 10&  -7.34&  1 \\ 
534.9465&  -0.310&  4&  -7.44&  12&  645.6875&   0.412& 10&  -6.61& 11 \\ 
551.2980&  -0.464&  7&  -7.00&  11&  820.1722&   0.300&  2&  -7.00& 11 \\ 
558.1965&  -0.555&  4&  -7.54&   9&  824.8796&   0.570&  2&  -7.00& 11 \\ 
558.8749&   0.358&  4&  -7.54&   9&  825.4730&  -0.400&  2&  -7.00& 11 \\ 
559.0114&  -0.571&  4&  -7.54&   9&  849.8023&  -1.416&  8&  -7.68&  9 \\ 
559.4462&   0.097&  4&  -7.54&   9&  854.2091&  -0.463&  8&  -7.68&  9 \\ 
585.7451&   0.240&  4&  -7.12&  11&  866.2141&  -0.723&  8&  -7.68&  9 \\ 
586.7562&  -1.570&  7&  -7.00&  11&  891.2068&   0.631& 10&  -7.21& 11 \\ 
610.2723&  -0.770&  4&  -7.19&   9&  892.7356&   0.808& 10&  -7.21& 11 \\ 
612.2217&  -0.319&  4&  -7.19&   9&  985.4759&  -0.228& 10&  -7.00& 11 \\ 
615.6023&  -2.497&  5&  -7.15&   9&  989.0628&   1.270& 10&  -7.01& 11 \\ 
616.1297&  -1.268&  4&  -7.15&   9&  993.1374&   0.051&  2&  -7.00& 11 \\ 
616.2173&  -0.090&  3&  -7.19&   9& 1183.8997&   0.290&  2&  -7.36& 11 \\ 
616.3755&  -1.286&  7&  -7.15&   9& 1194.9745&  -0.010&  2&  -7.36& 11 \\ 
\hline
\multicolumn{9}{l}{References}\\
\multicolumn{9}{l}{ 1:~VALD~~~~~~~~~~~~~~~~~~~~~~~~~~~~~~~~~~2:~Wiese et al. (1969)}\\ 
\multicolumn{9}{l}{ 3:~Wiese \& Martin (1980)~~~~~~~4:~Smith \& Raggett (1981)}\\
\multicolumn{9}{l}{ 5:~Kurucz \& Bell (1995)~~~~~~~~~6:~Morton (1991)}\\
\multicolumn{9}{l}{ 7:~Smith et al. (1988)~~~~~~~~~~~~~~~8:~Theodosiou (1989)}\\
\multicolumn{9}{l}{ 9:~Anstee \& O'Mara (1995), Barklem \& O'Mara (1997,1998)}\\
\multicolumn{9}{l}{ 10:~Opacity project~~~~~~~~~~~~~~~~~11:~fit of the Solar atlas}\\
\multicolumn{9}{l}{ 12:~Uns\"old  formula + fit of the Solar atlas}
\end{tabular}
\end{center}
\end{table}

$\bullet$  We retrieved the spectrum of Procyon from the UVES POP (Bagnulo et al., \cite{BJL03}). 
Procyon has a solar-like chemical composition, and its surface gravity, derived from  Hipparcos measurements (Perryman et al., \cite{PLK97}), is  log g=3.96.  We adopted a microturbulence $\xi_{t}=$1.8 \kms.
For this star, Mashonkina et al.  (\cite{MKP07}) adopting \Teff = 6510~K (Mashonkina et al., \cite{MGT03}) or \Teff = 6590~K  (Korn et al., \cite{KSG03}) 
found a subsolar abundance of Ca and a difference of about 0.2 dex between the NLTE calcium abundance  derived from the \Cau~ and \Cad~ lines.\\
 With \Teff=6510~K and our model atom, we derive for Procyon a near solar calcium abundance and a much better agreement between \Cau~ and \Cad:
 $\log \epsilon$(Ca)  = 6.25 $\pm 0.04$ from \Cau~lines and 6.27 $\pm 0.06$ from \Cad~lines
 (Fig. \ref{procyon}).\\

$\bullet$  We also tested our model on two classical metal-poor stars:
HD\,122563 (a star from our sample of giants) and HD\,140283.
 The spectrum of HD\,140283  was retrieved from the ESO POP (Bagnulo et al., \cite{BJL03}).
For HD\,122563 we adopted the parameters given in Table \ref{tabstars}, and for HD\,140283 the parameters given by Hosford et al. (\cite{HRG09}): \Teff=5750~K, \logg=3.4, $\xi_{\rm t}$=1.5~\kms. 
The agreement is good.  From the \Cau~lines we obtain $\rm\log \epsilon (Ca) =4.12 \pm 0.04$ and from the \Cad~lines $\rm\log \epsilon (Ca) =4.08 \pm 0.05$. In Fig. \ref{HD140283} we show the fit between the observed spectrum of HD\,140283 and the synthetic profiles computed with  $\rm\log \epsilon(Ca) =4.12$.

\begin{figure}[ht]
 \centering
  \includegraphics[width=0.24\textwidth,clip]{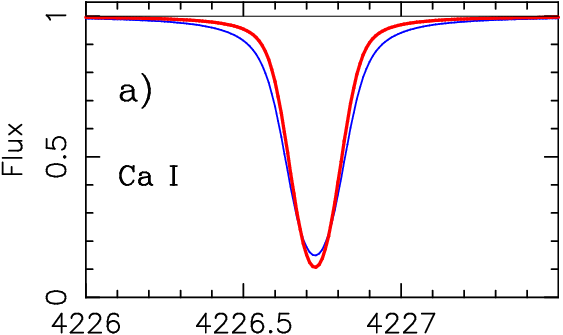}
  \includegraphics[width=0.24\textwidth,clip]{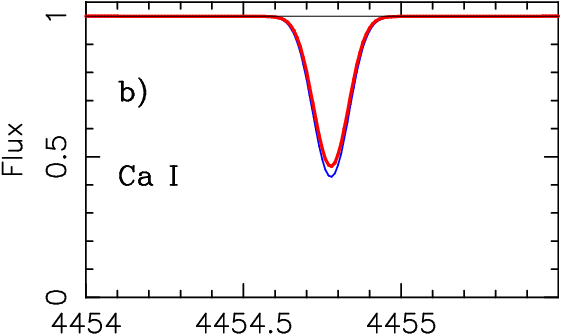}
  \includegraphics[width=0.24\textwidth,clip]{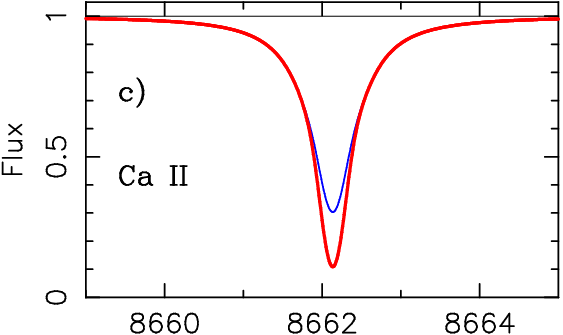}
   \caption[]{Profiles of three Ca lines computed for a giant star
   with $\rm [Ca/H] \approx -2.5$ with LTE (thin blue line) and NLTE
   (thick red line) hypotheses.  The wavelengths are given in \AA.\\
   a)
   The NLTE profile of the \Cau~ resonance line is narrower in the
   wings and deeper in the core.  b) In a \Cau~ subordinate line, for
   the same abundance of calcium the equivalent width computed under
   the NLTE hypothesis is slightly smaller.  c) The NLTE correction
   is important for the strong IR \Cad~ line (note that the scale in
   wavelength is different for this line), but the wings are not
   affected and a reliable calcium abundance can be deduced from these
   wings via LTE analysis.}
\label{spectra}
\end{figure}

\section{Calcium abundance in the EMP stars sample}  
\subsection{Comparison between the different systems}

In Fig.\ref{spectra} we show the influence of the NLTE effects on the profile of some typical calcium lines measurable in extremely metal-poor stars. Evidently, in particular the wings of the lines of the \Cad~infrared triplet are not sensitive to NLTE effects until the wings are strong enough.

\begin{figure*}[th]
 \centering
  \includegraphics[width=0.36\textwidth]{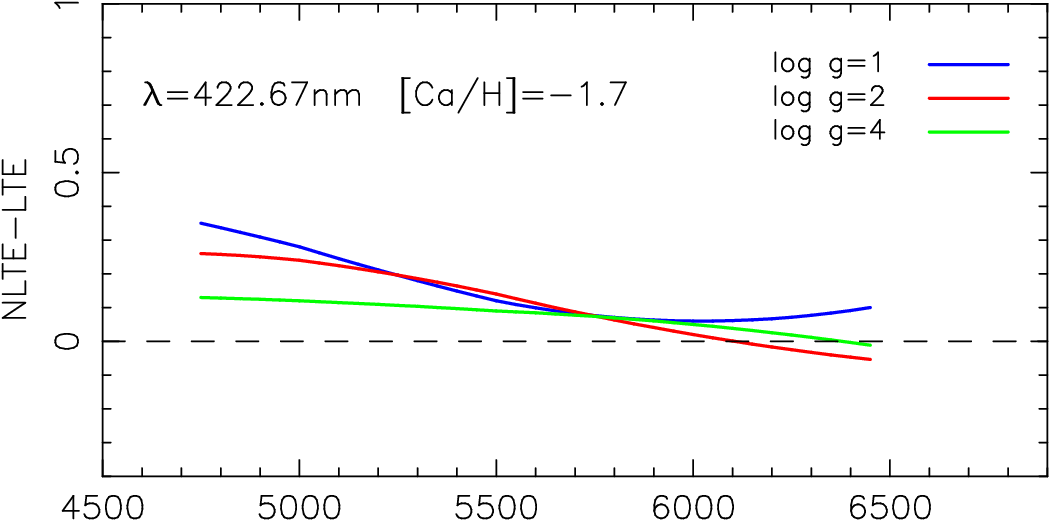}
  \includegraphics[width=0.36\textwidth]{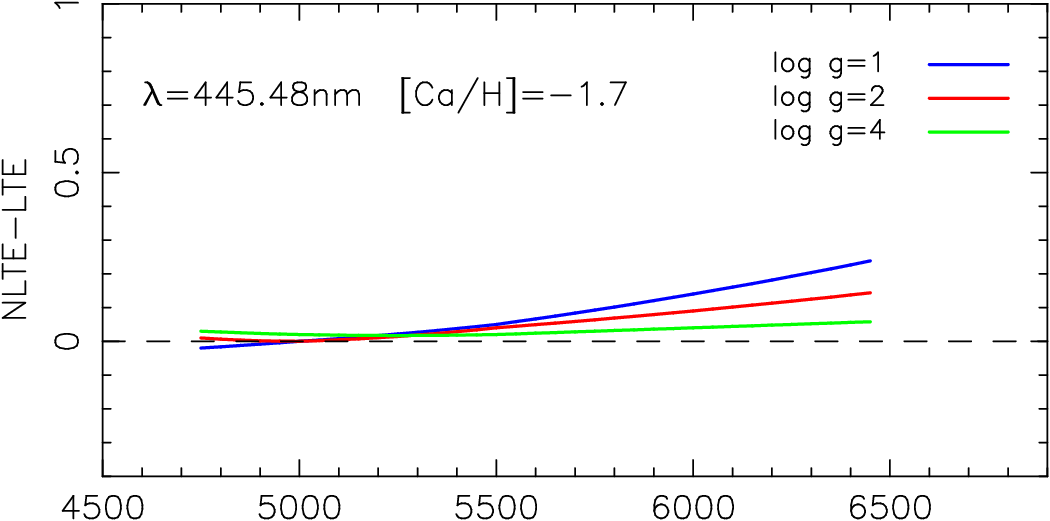}
  \includegraphics[width=0.36\textwidth]{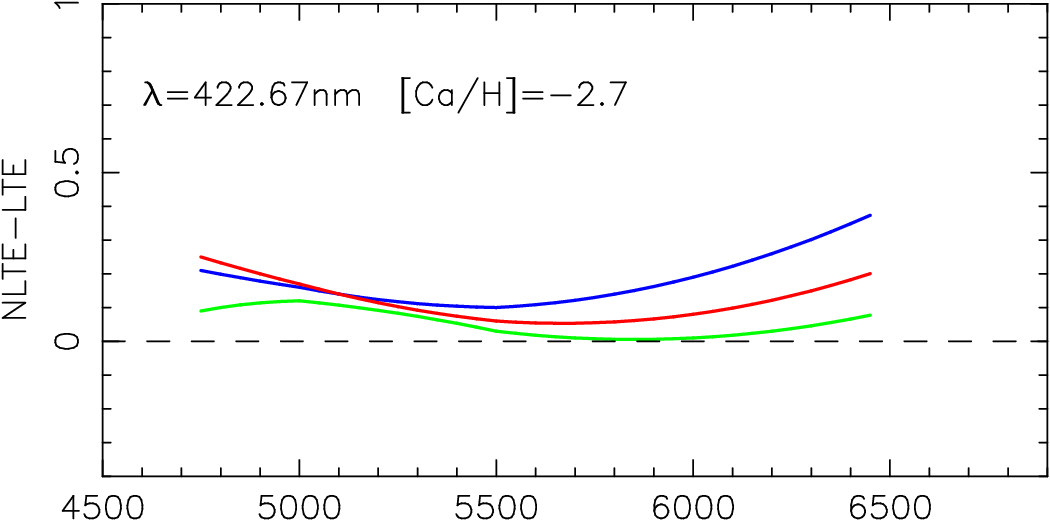}
  \includegraphics[width=0.36\textwidth]{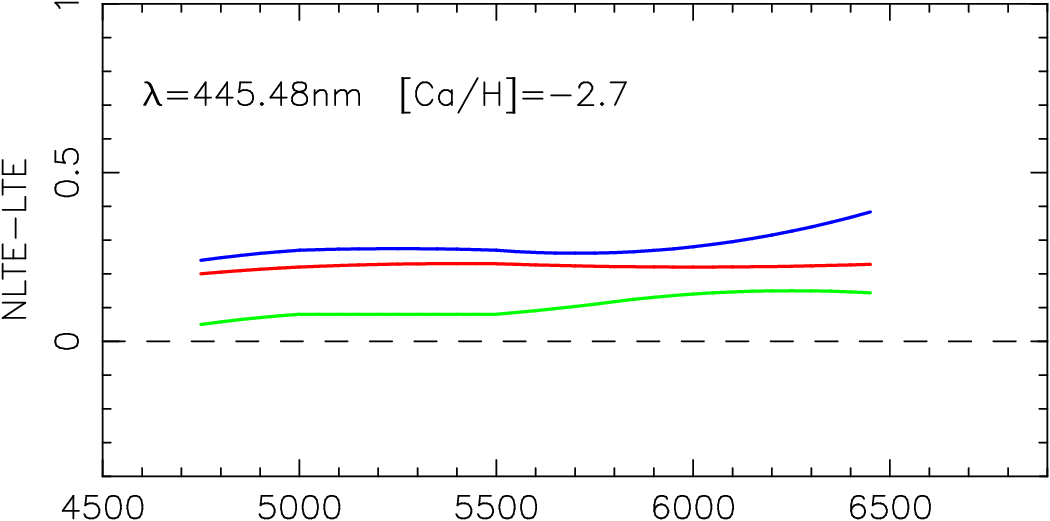}
  \includegraphics[width=0.36\textwidth]{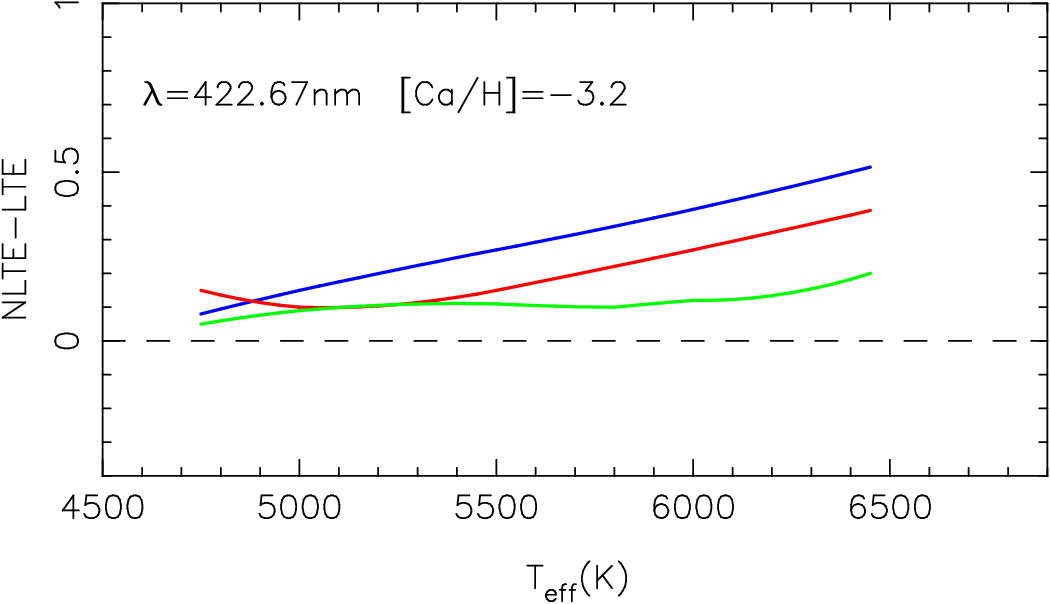}
  \includegraphics[width=0.36\textwidth]{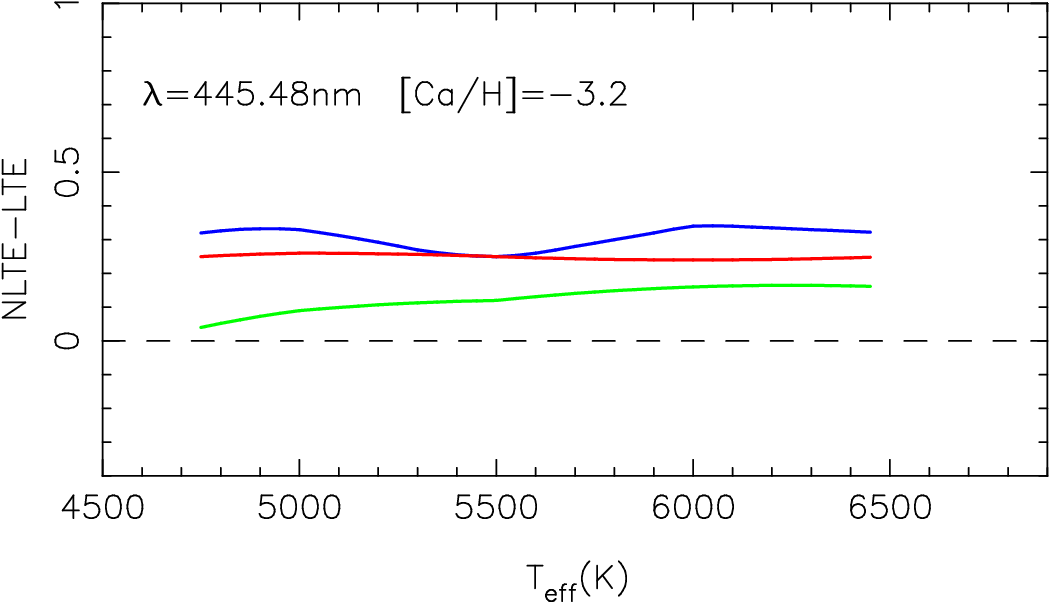}
\caption[]{NLTE corrections for the \Cau~resonance line at 422.67~nm
and for a typical subordinate line of \Cau~ at 445.48~nm.}
\label{corlte}
\end{figure*}

In Fig.  \ref{corlte} the
correction NLTE-LTE is given as a funtion of the temperature, the
gravity, and the calcium abundance for the \Cau~ resonance line and a
typical subordinate line of \Cau.  The NLTE corrections for different
temperatures, gravities, and abundances for all the Ca lines used in
this analysis (Table \ref{sunlist}) are available in electronic form\footnote
{http://cdsweb.u-strasbg.fr/cgi-bin/qcat?J/A+A/}.  
The influence of the
complex NLTE effects as a function of the stellar parameters and of the
characteristics of the lines has been discussed by Mashonkina et al.
(\cite{MKP07}) and Merle et al.  (\cite{MTP11}).

In Table \ref{tabstars} we give for each star the mean abundance
derived from the subordinate lines of \Cau~ (col. 6), the number of
lines used for the analysis (col. 7), and the internal error (col. 8).
Column 9 lists the abundance deduced from the resonance line of
\Cau, and  column 10 the abundance computed from the infra-red \Cad~
line at 866.21nm (only this line of the red \Cad~triplet is in the 
wavelength range of our spectra).  Columns 11 and 12 list 
[Ca/H] and [Ca/Fe] deduced from the subordinate lines of \Cau. 
For an easier comparison to the previous ``First Stars'' papers, the 
solar abundance of calcium was taken from Grevesse \& Sauval (\cite{GS00}) as in Cayrel et al. (\cite{CDS04}): $\rm log\epsilon_{(Ca)~\odot}=6.36$.
The oscillator strengths of the
calcium lines measured in our metal-poor stars were updated 
(to be the same as in section \ref{atomodel}). As a consequence,
for some lines (Table \ref {sunlist}) they are slightly different 
from the $\log~gf$ values used in Cayrel et al. (\cite{CDS04}) and 
Bonifacio et al. (\cite{BSC09}).  

{\bf The abundance of Ca was deduced from the equivalent widths
for the subordinate lines of \Cau}.  But for the resonance line, which is
often strong and slightly blended with Fe\,{\small I} and CH lines, a
fit of the profile was made.\\
{\bf The Ca abundance was generally deduced
from a fit of the wings (insensitive to NLTE effects)
for the red \Cad~ lines}.  However, in the
turnoff stars and in the most metal-poor giants, the wings almost
disappear (when the equivalent widths are less than 400~m\AA) 
and in this case, equivalent widths were used.

\begin{figure}[ht]
 \centering
  \includegraphics[width=0.45\textwidth]{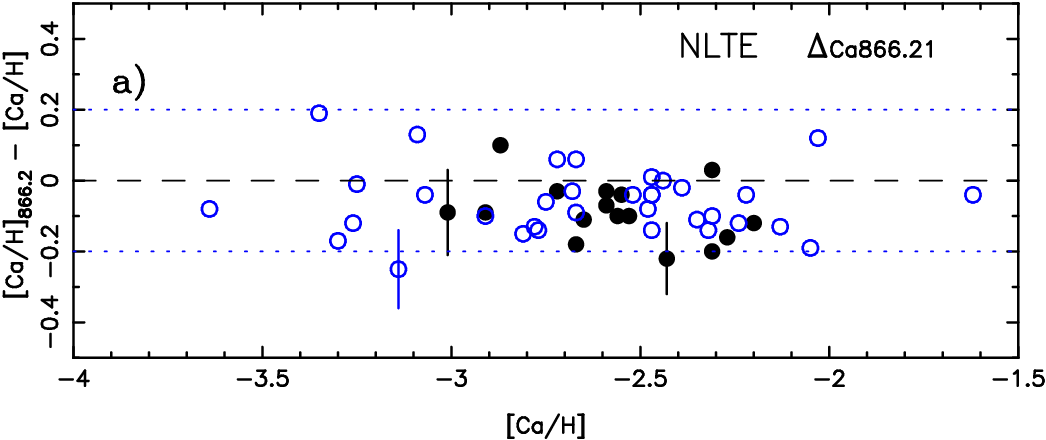}
  \includegraphics[width=0.45\textwidth]{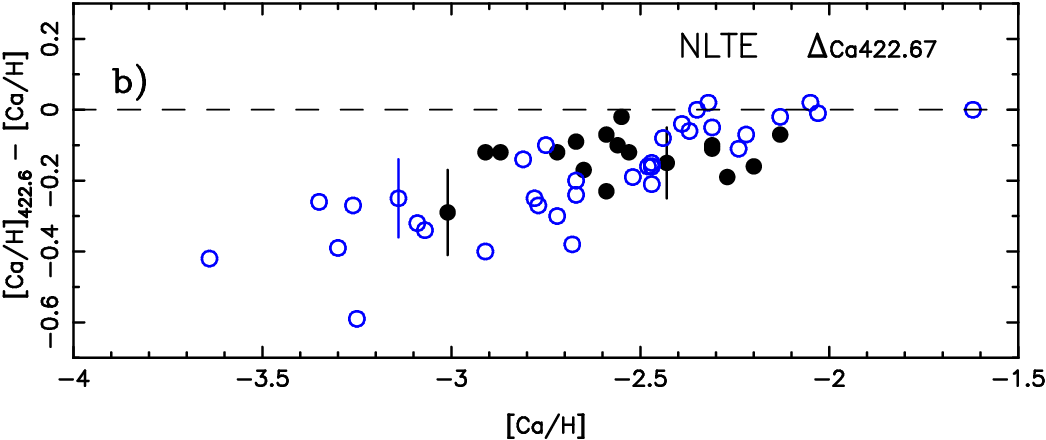}
\caption[]{Difference between [Ca/H] deduced from the
subordinate lines of \Cau~ and [Ca/H] deduced from
~~~a) the infrared triplet of \Cad~ and ~~~b) the resonance line of
\Cau.  The filled circles represent the turnoff stars and the open
symbols the giants. The error bar on [Ca/H], $\rm \Delta~Ca~866.21$
and $\rm \Delta~Ca~422.67$ is generally less than 0.1dex. It is indicated 
only when it exceeds 0.1dex.\\
a) The Ca abundance deduced from the line of the infrared triplet 
is, as a mean, 0.07 dex lower than the abundance derived from the 
subordinate lines of \Cau.\\
b) For $\rm [Ca/H]\approx -2$, the abundance 
deduced from the \Cau~ resonance line agrees quite well with the abundances deduced from
subordinate lines, but a discrepancy appears and
increases linearly when [Ca/H] decreases.  
It reaches about 0.4 dex for [Ca/H]=~--3.5.  }
\label{compAbCa}
\end{figure}

In Fig. \ref{compAbCa} we compare the abundances of calcium derived
from these different systems.  The abundance deduced from the line of
the infrared triplet of \Cad~ at 862.21nm is, as a mean, 0.07 dex
lower than the abundance deduced from the subordinate lines of \Cau~
(Fig.  \ref{compAbCa}a).  This small difference is quite
satisfactory since the abundance deduced from the \Cad~ lines depends
on the surface gravity of the model, and this gravity can be
affected by a systematic error since it has been derived from the
ionization equilibrium of iron under the LTE hypothesis (see Cayrel et
al., \cite{CDS04}).

\begin{figure}[ht]
 \centering
  \includegraphics[width=0.45\textwidth]{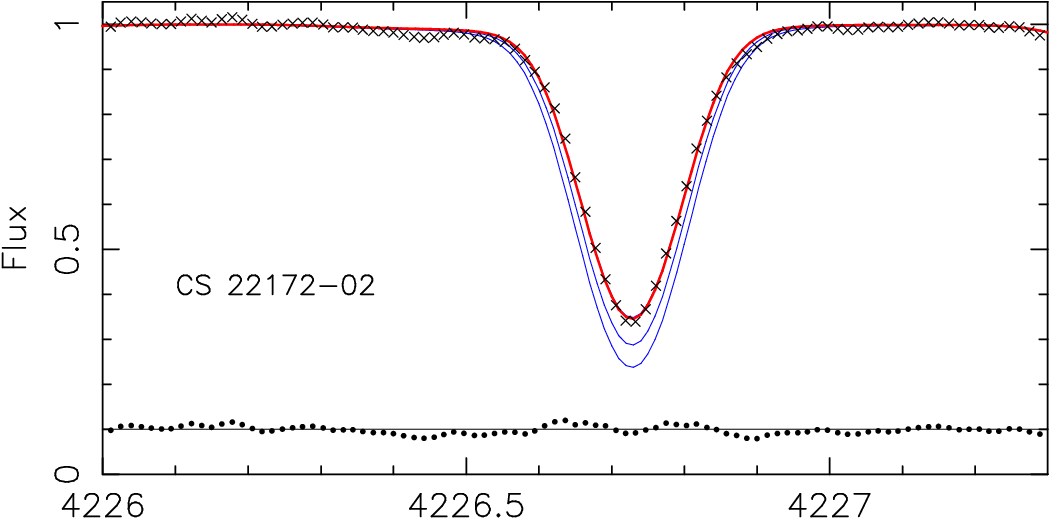}
\caption[]{Observed profile of the resonance line of \Cau~ (crosses) for
one typical metal-poor giant compared to theoretical profiles
computed with $\log~\epsilon \rm (Ca)=2.52$ (thick red line), 2.8 and 3.1 
(thin blue lines). The wavelengths are in \AA. 
The (very small) difference between the observed spectrum 
and the profile computed with $\log~\epsilon \rm (Ca)=2.52$ 
is shown at the bottom of the figure (dots, shifted by 0.1).  
For this star, the subordinate lines lead to $\log~\epsilon \rm (Ca)=3.11$ 
(Table \ref{tabstars}).
}
\label{prof4227}
\end{figure}

In Fig \ref{compAbCa}b, the calcium abundance derived from the
subordinate lines of \Cau~ is compared to the abundance deduced from
the \Cau~resonance line at 422.67nm.  It is well known that in very
metal-poor stars, under the LTE hypothesis, this resonance line leads 
to an underestimation of the calcium abundance: e.g. Magain (\cite{Mag88}),
Ryan et al.  (\cite{RNB96}).  But it was expected that NLTE
computations of the lines would remove this effect.  
For one typical metal-poor giant CS~22172-02 we show in Fig.
\ref{prof4227} the observed spectrum and theoretical spectra computed
with abundances $\log \epsilon$(Ca) = 2.52 (best fit), 2.8 and 3.1. 
The subordinate lines lead to a
Ca abundance of 3.11; this value is well established: seven of the
subordinate lines have an equivalent width {\bf stronger} than 9~m\AA~ and
the scatter in abundance is only 0.09~dex.

A similar discrepancy has been observed, after NLTE
computations, in several metal-poor stars by Mashonkina et al.
(\cite{MKP07}) and the authors suggested that the explanation could
lie with the 1D atmospheric models adopted for the computations, since the \Cau~ resonance line is formed over a more extended range of atmospheric depths than the subordinate lines.

\subsection{Influence of convection - 3D models}
 
\subsubsection{Shift of the calcium lines}
The atmospheres of cool stars are not static. Velocity and intensity
fluctuations caused by convection are 
observed in the Sun (Rutten et al., 2004) and in metal-poor stars.
Ram\'irez et al. (\cite{RCL10}) found that in HD\,122563 (a star of
our sample of giants) the cores of the \Feu~ lines are shifted
relative to the mean radial velocity of the star, this shift
increases with the equivalent width of the line. Weaker lines form
in deeper layers, where the granulation velocities and intensity
contrast are higher.
We also observe this phenomenon for the calcium lines in all
stars of our sample. The radial velocity of the stars were
determined from a constant set of iron lines, and relative to this
``zero point'', the radial velocity derived from the \Cau~ resonance 
line is about 0.4 \kms~ higher in the giants and 0.2 \kms~ higher 
in the dwarfs. In contrast, the radial velocity derived 
from the (weak) subordinate calcium lines is about 0.4 \kms~ lower 
in the giants and about 0.2 \kms~ lower in the dwarfs.
(We were able to measure precisely the shift of the subordinate lines only
on the ``blue spectra'' when these lines were larger than 20~m\AA.) 
At the resolution of our  VLT/UVES spectra with $R \approx
45000$, the asymmetry of the lines cannot be reliably measured.

The shift of the \Cau~ resonance line does not clearly depend on
[Ca/H], and therefore it does not seem that there is a clear correlation
between the shift of the Ca lines and the discrepancy between the
abundances deduced from the resonance or the subordinate lines.

\subsubsection{Abundance correction}
The largest abundance corrections caused by granulation effects occur at
low metallicities. This is mainly because the difference between the
1D and 3D predictions for the mean temperature of the outer layers of
metal-poor stars is very large (see e.g. Gonz\'alez Hern\'andez et
al., \cite{GBL10}).
We tried to investigate the change in abundances caused by thermal
inhomogeneities and differences in formation depth (3D corrections
hereafter) for the \Cau~resonance line at 422.67~nm and for a typical
subordinate line of \Cau~ at 445.48~nm. \\ 
--For a representative turnoff
star,  we used a 3D-CO5BOLD model (Freytag et al., \cite{FSD02}, 
\cite{FSL11}) from the CIFIST grid (see Ludwig et
al., \cite{LBS09}) with parameters (\Teff,  \logg,  [Fe/H]):
6270\,K/4.0/--3.0).\\ 
--For the giants we used two models
(4488\,K/2.0/--3.0 and 5020\,K/2.5/--3.0) from the CIFIST grid.  They
have both a gravity slightly higher than the ones in our sample of
giants, but no closer 3D model is available at the moment.

From these computations we found that for  [Fe/H]=--3,
[Ca/H]=--2.6, the 3D correction for the subordinate 
lines of \Cau~ in turnoff and in giant stars is very small.\\
In giants, the 3D correction seems
to be negligible also for the resonance line. 
But with the model of turnoff stars, we found  for
the resonance line a 3D correction of --0.44 dex, and therefore the 3D
correction increases the discrepancy between the subordinate lines and
the 422.67 resonance line. 
 
To date, it is not well understood why the abundance
of calcium deduced from the resonance line of \Cau~ is at very low
metallicity, lower than the abundance deduced from the subordinate
lines. 
It would be interesting to repeat the 3D computations with more metal-poor 
models (not available today).
A fully consistent 3D NLTE model atmosphere and line formation
scheme is currently beyond our reach.


\section{Results and discussion}

\begin{figure}[ht]
 \centering
  \includegraphics[width=0.45\textwidth]{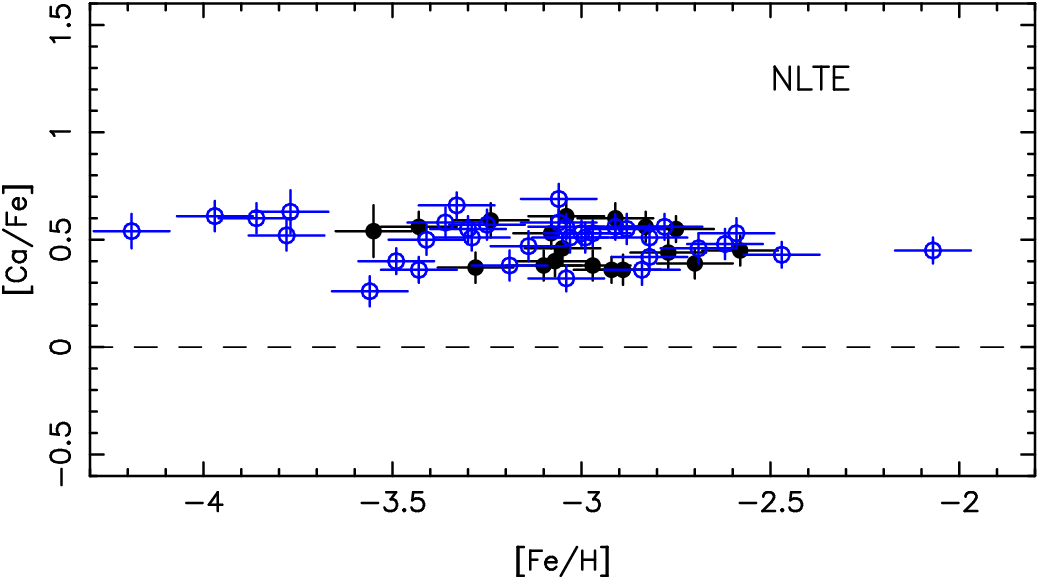}
\caption[]{ [Ca/Fe] vs.  [Fe/H] in our sample of EMP stars.  Symbols
as in Fig \ref{compAbCa}.  Turnoff and giant stars agrees quite well, 
[Ca/Fe] is constant in the range
$\rm-4.3<[Fe/H]<-2.5$ and the mean value is equal to [Ca/Fe]=0.5, a
value slightly higher than the LTE value [Ca/Fe]=0.35 (see Bonifacio
et al., \cite{BSC09}).  }
\label{cafe}
\end{figure}

In this section we adopt the Ca abundance deduced from the
subordinate lines of \Cau~ (Tab.  \ref{tabstars}, columns 11 and 12).

In Fig.  \ref{cafe} we present the new relation between [Ca/Fe] and
[Fe/H].  The error bar plotted in for the subordinate lines the figure is the quadratic sum of the
error due to the uncertainty of the model and the random error of the
Ca abundance derived from the subordinate lines.  The agreement
between the turnoff and the giant stars is excellent, the ratio
[Ca/Fe] is constant in the interval $\rm-4.5<[Fe/H]<-2.5$.  The slope
of the regression line is --0.05.  The scatter of [Ca/Fe] is a little
smaller when NLTE effects are taken into account: 0.09 from NLTE
computations compared to 0.10 from LTE computation .

\begin {figure}[ht]
\begin {center}
\resizebox  {4.2cm}{2.4cm}
{\includegraphics {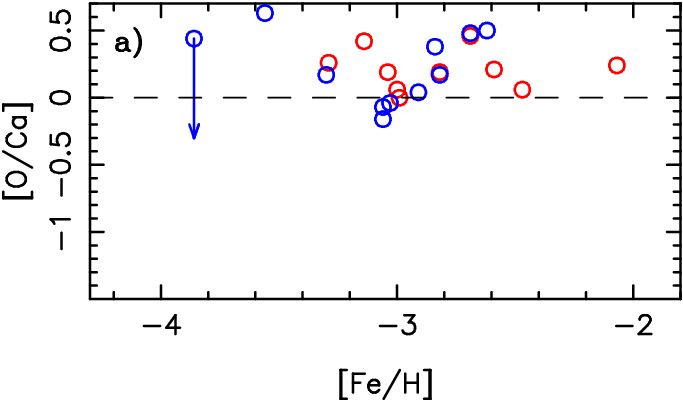} }
\resizebox  {4.2cm}{2.4cm}
{\includegraphics {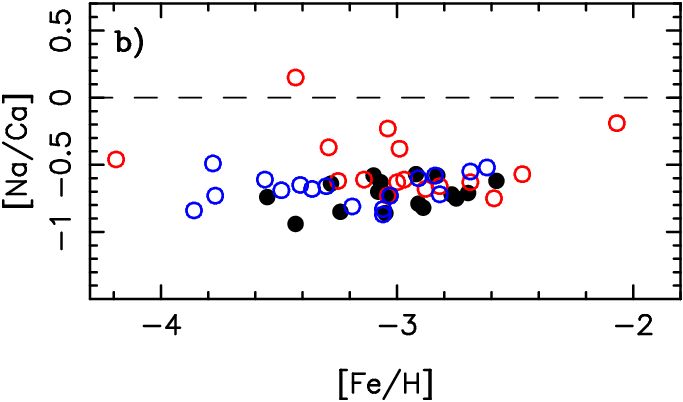} }
\resizebox  {4.2cm}{2.4cm}
{\includegraphics {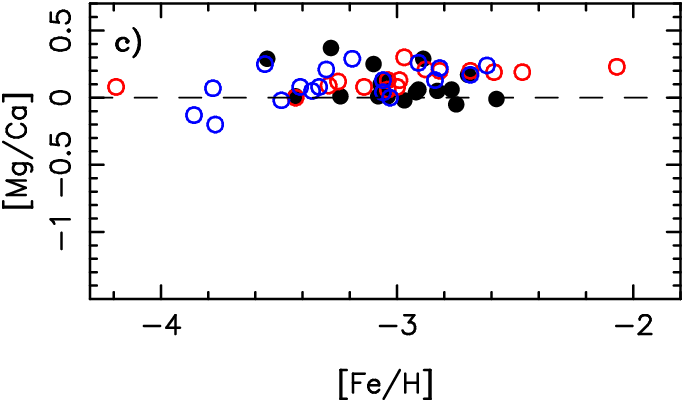} }
\resizebox  {4.2cm}{2.4cm}
{\includegraphics {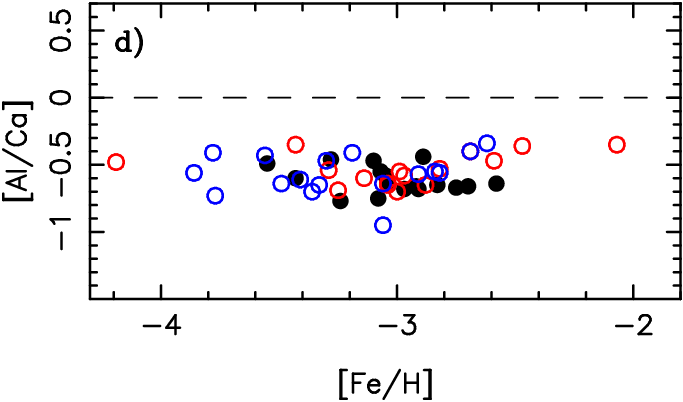} }
\resizebox  {4.2cm}{2.4cm}
{\includegraphics {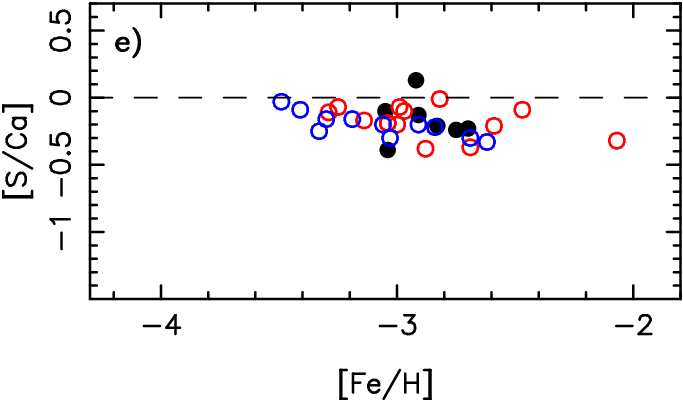} }
\resizebox  {4.2cm}{2.4cm}
{\includegraphics {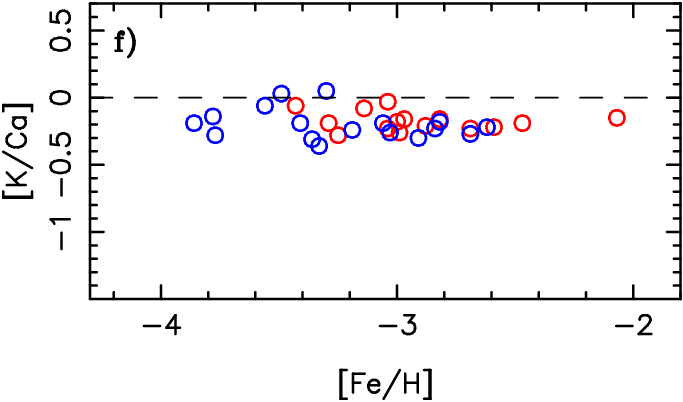} }
\caption{Abundance ratios of O, Na, Mg, Al, S, and K relative to Ca 
in the early Galaxy.  
The abundances of all these elements were computed
taking into account the NLTE effect.  The black dots represent the
turnoff stars, the open circles the giants (blue for the unmixed
giants and red for the mixed giants).} \label {el-ca} \end {center}
\end {figure}

In Fig.  \ref{el-ca} we present the behavior of the abundances of
O, Na, Mg, Al, S and K relative to Ca.  The computation of these 
abundances take into account the NLTE effects (Andrievsky
et al.  \cite{ASK07}, \cite{ASK08}, \cite{ASK10} and Spite et al.
\cite{SCA11}).  Since the abundance of oxygen has been determined from
the forbidden oxygen line at 630~nm (Cayrel et al.  \cite{CDS04}), it
is free of NLTE effects.

In Fig. \ref{el-ca}, different symbols are used for mixed and unmixed
giants.  After Spite et al.  (\cite{SCP05}, \cite{SCH06}) we call
``mixed giants'', those where, ownng to mixing with deep layers, carbon has been
partially transformed into nitrogen ($\rm[C/N]<-0.6$), 
part of $\rm ^{12}C$ has been transformed into
$\rm ^{13}C$ ($\rm^{12}C/^{13}C < 10$), 
and lithium is not detected (lithium has
been severely depleted by this mixing).  In the HR
diagram, these ``mixed giants'' are located
above the ``bump''.  It has been found (Andrievsky et al.,
\cite{ASK07}) that some mixed giants are enriched in sodium, this is
visible in Fig.  \ref{el-ca}b.  This Na-enhancement reflects a deep
internal mixing (or an AGB status, or a contamination by AGB stars), but
it does not reflect an anomaly of the chemical composition of the
cloud that formed the star.  Therefore the mixed stars cannot be used
to determine the ratio [Na/Ca] in the early galactic matter.

\begin {table}[ht] \caption {Mean value of the ratios [X/Fe], [X/Ca]
and [X/Mg] in the interval $\rm -3.6 < [Fe/H] < -2.5$ and scatter
around the mean} \label {dispers}
\begin{center}
\begin{tabular}{l c c c c c c c c c}
\hline
\\
X &$\rm \overline{[X/Fe]}$&$\sigma_{Fe}$&$\rm
\overline{[X/Mg]}$&$\sigma_{Mg}$&$\rm
\overline{[X/Ca]}$&$\sigma_{Ca}$\\
\hline
O  &   0.67 & 0.17   &     0.07 & 0.20    &   0.23 & 0.22   \\
Na &  -0.22 & 0.11   &    -0.82 & 0.14    &  -0.70 & 0.11   \\
Mg &   0.68 & 0.13   &     --   & --      &   0.12 & 0.11   \\
Al &  -0.10 & 0.11   &    -0.71 & 0.11    &  -0.58 & 0.13   \\
S  &   0.31 & 0.12   &    -0.31 & 0.15    &  -0.19 & 0.12   \\
K  &   0.31 & 0.09   &    -0.32 & 0.15    &  -0.18 & 0.10   \\
Ca &   0.49 & 0.10   &    -0.12 & 0.11    &   --   & --     \\
\\
\hline
\end {tabular}  
\end {center}  
\end {table}

Since the abundance ratios of O, Na, Al, S, and K relative to Ca are
fairly flat vs. [Fe/H] in the central part of the diagram, 
we can define a mean value in this central interval, say $\rm -3.6 < [Fe/H] < -2.5$ as
has been done with Mg in Andrievsky et al.  (\cite{ASK10}).  The mean
values of [X/Fe], [X/Mg] and [X/Ca] are given in Table \ref{dispers}
with the corresponding scatter. However, an anti-correlation seems to exist between [S/Ca] and [Fe/H] and also [S/Mg] and [Fe/H]: the Kendall $\tau$ coefficient is 98.9\% for [S/Ca]  and 99.1\% for [S/Mg]. In these computations the weight of BD~+$17^{\circ}3248$ ([Fe/H]=--2.2) is important. This star is quite peculiar: according to For \& Sneden (\cite {FS10}) it is a red horizontal branch star strongly enriched in heavy elements (Cowan et al., \cite{CSB02}). However, if this star is removed from the computations, the Kendall $\tau$ coefficient remains high: 97.6\% for [S/Ca]  and 97.9\% for [S/Mg].

One turnoff star, BS~16023-046, and one unmixed giant CS~22956-50,
have abnormally strong and broad sodium D lines. Their radial velocity (--7.5\,\kms ~and --0.1\,\kms) is low, and the stellar
lines are very probably blended with interstellar lines. These stars have
not been plotted on Fig. \ref{el-ca}b and were not taken into account in
computing the mean value and the scatter of [Na/Fe], [Na/Mg] 
and [Na/Ca] in Table \ref{dispers}.

It is interesting to take advantage of the high quality of the data to
compare in Table \ref{dispers} the scatter around the mean value when
Fe, Mg, or Ca are taken as reference elements.  The scatter of [O/Fe],
[O/Mg] or [O/Ca] is fairly large: about 0.2~dex but this can be
explained by the scatter of the oxygen abundance due to the reduced 
number of oxygen lines and the difficulty of
the measurements.  For the other ratios the scatter lies between 0.09
and 0.15~dex and we consider that this difference of 0.06 dex is significant.  
The scatter is always significantly smaller when Ca is taken as the
reference element, with two exceptions:\\
$\bullet$ Oxygen: but in this case we have seen that the scatter is dominated by the error on the measurement of the very weak oxygen line.\\
$\bullet$ Aluminum: Al is better correlated with Mg than with Ca.  But
this correlation (opposite to the well known anti-correlation found
in globular cluster stars)
has at least one exception: CS~29516-024 is
rather Al-poor but it is (relatively) Ca-rich and Mg-rich, and it does
not seem possible to explain these differences by errors of
measurements.  A similar correlation between [Al/Fe] and [Mg/Fe] has
been also suggested by Suda et al. (\cite{SYK11}) even in a large 
but very inhomogeneous sample of metal-poor stars.

\section{Concluding remarks}

We determined the calcium abundance in a
homogeneous sample of of 53 metal-poor stars (31 of them with
$\rm[Fe/H]<-2.9$) taking into account departures from LTE. We have
shown that the trend of the ratio [Ca/Fe] vs. [Fe/H], below [Fe/H]=--2.7 is almost flat and derived a new mean value of [Ca/Fe] in
the early Galaxy: $\rm[Ca/Fe]=0.5 \pm 0.09$.

Generally speaking, our NLTE calculations agree quite well with those of Mashonkina et al.  (\cite{MKP07}). 
However, Mashonkina et al. had found different Ca abundances from the \Cau~ and \Cad~ lines in Procyon.  
With our new model atom of calcium, we are able to reconcile the calcium abundance deduced from neutral and ionized calcium lines in Procyon.  We derived 
 $\log \epsilon$(Ca)  = 6.25 $\pm 0.04$ from \Cau~lines and 6.27 $\pm 0.06$ from \Cad~lines (6.33 and 6.40 under the LTE hypothesis).
Moreover, the calcium abundance is found to be solar, as expected. 

In metal-poor stars (below [Ca/H]=--2.5), a
discrepancy clearly appears between the Ca abundances deduced 
from either the resonance \Cau~ line, or the \Cau~ 
subordinate lines. A rough estimation of the effect of convection
on the profile of these lines does not explain this discrepancy.

In the stars of our sample (giants and turnoff stars) the wavelengths of the calcium lines are shifted by convection as a function of the equivalent width of the lines as has been found for the iron lines by Ram\'{\i}rez et al. (\cite{RCL10}) in HD\,122563.

The scatter around the mean value of [Ca/Fe] is small but since the
NLTE correction is about the same for all the subordinate lines used
in this analysis, the scatter is almost the same as it was in the LTE analyses (Cayrel et al., \cite{CDS04}, Bonifacio et al., \cite{BSC09}).

The abundance of the light metals O, Na, Al, S, K is well
correlated with the calcium abundance.  The correlation is generally a
little better than it is with the magnesium abundance (with exception
of the tight aluminium/magnesium correlation).  However, it is striking
that in Table \ref{dispers}, even when NLTE is taken into account, the
correlation of the light elements O, Na, Al, S, K with Ca or Mg
(all supposed to be formed mainly by hydrostatic fusion 
of C, Ne or O), is never better than the correlation with iron,
providing some support to supernovae models predicting nucleosynthesis of all these elements predominantly in explosive modes (e.g. Nomoto et al. \cite{NTU06}).


\begin {acknowledgements} The authors thank the referee,   who indicated new $\log gf$ values for the high excitation \Cad~ lines. 
This work has been supported in part by the
"Programme National de Physique Stellaire" (CNRS).  
S.A. kindly thanks the Observatoire de Paris, the CNRS, and the laboratory GEPI for their hospitality and support during his stay in Meudon.
He acknowledges the National Academy of Sciences of Ukraine and the franco-Ukrainian exchange program  for its financial support under contract UKR CDIV N24008.
\end {acknowledgements}


\begin{thebibliography}{}


\bibitem[1973]{All73}
Allen C.W., 1973, Astrophysical Quantities, Athlone Press, London

\bibitem[1996]{AAM96}   
Alonso A., Arribas S., Mart\'{i}nez-Roger C., 1996, A\&A 313, 873

\bibitem[1999]{AAM99}   
Alonso A., Arribas S., Mart\'{i}nez-Roger C., 1999, A\&AS 140, 261

\bibitem[2001]{AAM01}   
Alonso A., Arribas S., Mart\'{i}nez-Roger C., 2001, A\&AS 376, 1039

\bibitem[2007]{ASK07}   
Andrievsky S., Spite M., Korotin S. et al., 2007, A\&A 464, 1081

\bibitem[2008]{ASK08}   
Andrievsky S. M., Spite M., Korotin S. A., Spite F., Bonifacio P.,
Cayrel R., Hill V., Fran\c cois P., 2008, A\&A 481, 481

\bibitem[2010]{ASK10}   
Andrievsky S., Spite M., Korotin S. et al., 2010, A\&A 509, 88

\bibitem[1995]{AO95} 
Anstee S. D.,  O'Mara B. J., 1995, MNRAS, 276, 859

\bibitem[2009]{AGS09} 
Asplund M. et al., Grevesse N., Sauval J., Scott P., 2009, ARA\&A,
47, 481 


\bibitem[2003]{BJL03}
Bagnulo S., Jehin E., Ledoux C. et al., 2003, The Messenger, 114, 10

\bibitem[1997]{BO97}
Barklem P. S., O'Mara B. J., 1997, MNRAS 290, 102

\bibitem[1998]{BO98}
Barklem P. S., O'Mara B. J., 1998, MNRAS 300, 863

\bibitem[1998]{BOR98}
Barklem P. S., O'Mara B. J., Ross, J. E., 1998, MNRAS 296, 1057

\bibitem[2007]{BMS07}
Bonifacio P., Molaro P., Sivarani T. et al., 2007, A\&A 462, 851

\bibitem[2009]{BSC09}
Bonifacio P., Spite M., Cayrel R. et al., 2009, A\&A 501, 519

\bibitem[1995]{BCT95}
Burgess,A., Chidichimo M. C., Tully J. A., 1995, A\&A 300, 627

\bibitem[2007]{CL07}
Caffau E., Ludwig H.-G., 2007, A\&A 467, L11

\bibitem[1986]{Car86}
Carlsson M., 1986, Uppsala Obs. Rep. 33


\bibitem[2004]{CDS04}
Cayrel R., Depagne E., Spite M. et al., 2004, A\&A 416, 1117

\bibitem[2002]{CSB02}
Cowan J.J., Sneden C., Burles S., Ivans I.I., Beers T.C., Truran J.W. et al., 2002, ApJ 572, 861

\bibitem[2000]{DDK00}
Dekker H., D'Odorico S., Kaufer A., Delabre B., Kotzlowski H., 2000, Proc. SPIE 4008, 534

\bibitem[2010]{FS10}
For B.-Q., Sneden C., 2010, AJ 140, 1694

\bibitem[2002]{FSD02}
Freytag B., Steffen M., Dorch B., 2002, AN 323, 213

\bibitem[2011]{FSL11}
Freytag B., Steffen M., Ludwig H.-G., Wedemeyer-B\"ohm S.,
Schaffenberger W., Steiner O., 2011, JCoPh, 231, 919

\bibitem[2010]{GBL10}
Gonz\'alez Hern\'andez J.I., Bonifacio P., Ludwig H.-G., Caffau E.,
Behara N. T., Freytag B., 2010, A\&A 519, 46

\bibitem[2000]{GS00}
Grevesse N., Sauval A.J., 2000, "Origin of the elements  in the solar 
system. Implications of post-1957 Observations", ed. O. Manuel, 
Kluwer Academic/Plenum Publishers, p.261

\bibitem[1995]{HH95}
Hirata R., Horaguchi T., 1995, Catalogue of Atomic Spectroscopic
Lines, Vol. 6 (Strasbourg : CDS), 69

\bibitem[2009]{HRG09}
Hosford A., Ryan S. G., Garc'a PŽrez A. E., Norris J. E., Olive K. A.,
2009, A\&A 493, 601

\bibitem[2002]{ISS02}
Ivanova D. V., Sakhibullin N. A., Shimanskii V. V., 2002, ARep 46, 390

\bibitem[2006]{KUN06}
Kobayashi Ch., Umeda H., Nomoto K. et al., 2006,
ApJ 653, 1145

\bibitem[2003]{KSG03}
Korn, A., Shi, J., Gehren, T., 2003, A\&A, 407, 691

\bibitem[2008]{Kor08}
Korotin S.A., 2008, Odessa Astronomical Pub. 21, 42

\bibitem[1999]{KAL99}
Korotin S.A., Andrievsky S.M., Luck R.E., 1999, A\&A 351, 168

\bibitem[1992]{Kur92}
Kurucz R.L., 1992, RMxAA 23, 181

\bibitem[1993]{Kur93}
Kurucz R.L., 1993, SAO, Cambridge, CDROM18 

\bibitem[1995]{KB95}
Kurucz R.L., Bell B., 1995, SAO, Cambridge, CDROM23 

\bibitem[1984]{KFB84}
Kurucz R.L., Furenlid I., Brault J., Testerman L., 1984, National Solar Observatory Atlas, Sunspot, New Mexico: National Solar Observatory, 1984  

\bibitem[2009]{LPG09}
Lodders K., Plame H., Gail H.-P., 2009, in Landolt-B\"ornstein, New Series,  Volume VI/4B Chapter 4.4. Edited by J.E. Tr\"umper, Berlin Heifelberg New York: Springer-Verlag p. 560

\bibitem[2009]{LBS09}
Ludwig H.-G., Behara N. T., Steffen M., Bonifacio P., 2009, A\&A 502, 1L

\bibitem[1988]{Mag88}
Magain P., 1988, in IAU Symp 132, ``The impact of Very High S/N
spectroscopy on Stellar Physics'', G. Cayrel de Strobel \& M. Spite
eds (Dordrecht, Kluwer) p.485

\bibitem[2003]{MGT03}
Mashonkina L., Gehren T., Travaglio C., Borkova T., 2003, A\&A, 397, 275

\bibitem[2007]{MKP07}
Mashonkina L., Korn A.J., Przybilla N., 2007, A\&A 461, 261 

\bibitem[2007]{MBB07}
Mel\'endez M., Bautista M.A., Badnell N.R., 2007, A\&A 469, 1203 

\bibitem[2011]{MTP11}
Merle T., Th\'evenin F., Pichon B., Bigot L., 2011, MNRAS 418, 863

\bibitem[1991]{Mor91}
Morton D.C., 1991, ApJS 77,119 

\bibitem[2006]{NTU06}
Nomoto K., Tominaga N., Umeda H. et al., 2006, Nucl. Phys. A 777, 424

\bibitem[1997]{PLK97}
Perryman M. A. C., Lindegren L., Kovalevsky J. et al., 1997, A\&A, 323, L49


\bibitem[2010]{RCL10}
Ram\'{\i}rez I., Collet R., Lambert D.L., Allende Prieto C., Asplund
M., 2010, ApJL 725, L223

\bibitem[2004]{RHB04}
Rutten R. J., Hammerschlag R. H., Bettonvil F. C. M., S\"{u}tterlin
P., \& de Wijn A. G., 2004, A\&A  413, 1183

\bibitem[1996]{RNB96}
Ryan S.G., Norris J.E., Beers T.C., 1996, ApJ 471, 1996

\bibitem[2001]{SB01}
Samson A. M., Berrington K. A., 2001, ADNDT 77, 87

\bibitem[1962]{Sea62}
Seaton M.J., 1962, in Atomic and Molecular processes (New York:
Academic Press)

\bibitem[1988]{Smi88}
Smith G., 1988, J. Phys. B, 21, 2827

\bibitem[1981]{SR81}
Smith G., Raggett D. St. J., 1981, J. Phys. B, 14, 4015

\bibitem[2005]{SCP05}
Spite M., Cayrel R., Plez B. et al., 2005, A\&A 430, 655 

\bibitem[2006]{SCH06}
Spite M., Cayrel R.,  Hill V. et al., 2006, A\&A 455, 291     

\bibitem[2011]{SCA11}   
Spite M., Caffau E.,  Andrievsky S. et al., 2011, A\&A 528, 9

\bibitem[1984]{SH84}
Steenbock W., Holweger H., 1984,  A\&A 130, 319

\bibitem[2011]{SYK11}
Suda T., Yamada S., Katsuta Y., Komiya Y. et al., 2011, MNRAS 412, 843

\bibitem[1985]{SC85}
Sugar J., Corliss C., 1985, J. Phys. Chem. Ref. Data, 14, Suppl. No. 2

\bibitem[1989]{Theo89}
Theodosiou C. E., 1989, Phys. Rev. A, 39, 4880

\bibitem[1962]{VR62}
Van Regemorter H., 1962, ApJ 136, 906

\bibitem[1981]{VAL81}
Vernazza J. E., Avrett E. H., Loeser R., 1981, ApJS, 45, 635

\bibitem[1980]{WM80}
Wiese W.L., Martin G.A., 1980, NSRDS-NBS 68, Part II 

\bibitem[1969]{WSM69}
Wiese W.L. Smith M. W., Miles B.M., 1969, NSRDS-NBS 22, vol.2 


\end{thebibliography}
\end{document}